\documentclass[prl,superscriptaddress,twocolumn,showpacs]{revtex4}
\usepackage{amsfonts,amssymb,amsbsy,bm,graphicx,amsmath,times,color}

\begin{document}

\title{Quantum reconstruction of the mutual coherence function}

\author{Z. Hradil}
\affiliation{Department of Optics, Palacky University,
17. listopadu 12, 771 46 Olomouc, Czech Republic}

\author{J. \v{R}eh\'{a}\v{c}ek}
\affiliation{Department of Optics, Palacky University,
17. listopadu 12, 771 46 Olomouc, Czech Republic}

\author{L. L. S\'{a}nchez-Soto}
\affiliation{Departamento de \'Optica, Facultad de F\'{\i}sica,
Universidad Complutense, 28040~Madrid, Spain}

\begin{abstract}
  Light is a major carrier of information about the world around us,
  from the microcosmos to the macrocosmos.  The present methods of
  detection are sensitive both to robust features, such as intensity,
  or polarization, and to more subtle effects, such as correlations.
  Here we show how wave front detection, which allows for registering
  the direction of the incoming wave flux at a given position, can
  be used to reconstruct the mutual coherence function when combined
  with some techniques previously developed for quantum information
  processing.
\end{abstract}

\pacs{03.65.Wj, 03.65.Ta, 42.50.Tx}

\maketitle

\textit{Introduction.---}
Three-dimensional objects emit characteristic wave fronts that are
determined by different object features, such as shape, refractive
index, density, or temperature. Wave front sensing constitutes thus an
invaluable tool to acquire information about our surroundings,
especially because ordinary detection devices are not phase sensitive
and only capture two-dimensional intensity data.

The measurement of the wave front phase distribution is a key issue
for many applications, such as noncontact metrology, adaptive optics,
high-power laser systems, and
ophthalmology~\cite{Tyson:1998,Roddier:1999}. Since the frequency of
light waves is so high, no detectors of sufficient time resolution
exist and indirect methods for phase measurements have been
developed~\cite{Geary:1995,Buse:2000}, each one with their own pros
and cons. In short, practical and robust wave front sensing is still
an unresolved and demanding problem.

The purpose of this Letter is to point out that these standard methods
may be underrated and do not fully exploit the potential of registered
data. By recasting their functioning principle in a quantum language,
one can immediately foresee that, by using the methods of tomographic
reconstruction~\cite{Rehacek:2004}, these devices allow in fact for a
full evaluation of the mutual coherence function of the signal, which
conveys full information. This should be compared with the partial
phase information retrieved by the standard wave front reconstruction
techniques, where full coherence of the detected signal is assumed and
imperfect correlations are ignored. Going beyond such a standard 
interpretation may constitute a substantial step ahead as it is indeed an
interdisciplinary task involving wave and statistical optics, as well
as protocols of quantum state reconstruction.

\begin{figure}[b]
   \centerline{\includegraphics[height=6cm]{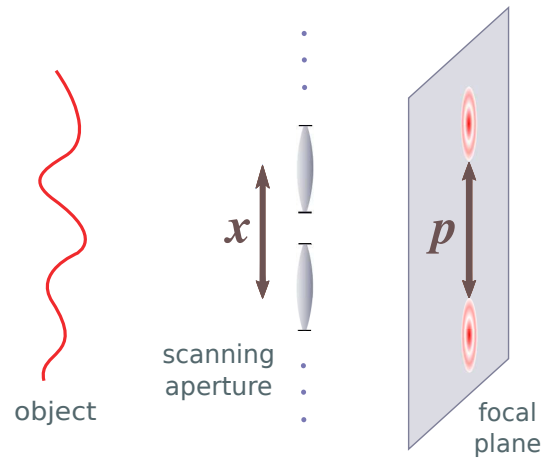}}
\caption{Simplified scheme of a general scanning setup.
Two of the microlenses of the periodic array are shown. 
By moving the aperture, the field is sampled at different positions,
whereas by moving the detector in the focal plane  
the field is sampled at different angles.
\label{fig:scan}}
\end{figure}

\textit{Classical theory of  wave front measurements.---}
To be specific, we sketch the principle of a Hartmann-Shack
sensor~\cite{Platt:2001}, which is general enough for our purposes
here and is schematized in Fig.~\ref{fig:scan}. To keep the discussion
as simple as possible, we restrict ourselves to a
one-dimensional model and denote by $x$ the position in the scanning
aperture.  The incoming light field is divided into a number of
subapertures by a microlens array that creates focal spots.  The
deviation of the spot pattern from a reference measurement allows the
local direction angles to be derived, which in turn allows the
reconstruction of the wave front. In addition, the light intensity
distribution at the detector plane can be obtained by integration and
interpolation between the foci.

These devices provide a simultaneous detection of position $x$ and
angular spectrum $p$ of the incident radiation, which is determined by
the position of the detected signal on the screen. As we shall see
below, $p$ determines directly the transverse impulse $p_{x}$ of
incident radiation. Let $\Phi$ be the complex amplitude of the signal
emitted by the source. The propagation of this signal is described by
the convolution with the optical transfer function
$h$~\cite{Goodman:2005}, so that at the plane of the apertures it can
be expressed as
\begin{equation}
  \label{prop}
  \Phi_{\mathrm{ap}} ( x )  = \int dx^{\prime} \,
  \Phi ( x^{\prime} ) \, h (x^{\prime} - x ) \, .
\end{equation}
The amplitude in the focal plane due to the $j$th microlens of
extension  $\mathcal{A}_{j}$  thus reads
\begin{equation}
  \label{coherent amplitude}
  \Phi_{j} ( p )  = \int_{\mathcal{A}_{j}}
  dx \,
  \Phi_{\mathrm{ap}} ( x )
  A_{j} ( x )
  \exp \left ( i \frac{k}{f} x  p \right ) \, ,
\end{equation}
where $k = 2\pi/\lambda $ is the wave number, $f$ the focal length,
and the integration is performed in the transverse plane.
The function $A_{j}$  is the aperture (or pupil) function of the
$j$th microlens. Since only intensities can be detected, the
corresponding signal at the focal plane is
\begin{eqnarray}
  \label{projection}
  S_{j,p} \equiv \langle |\Phi_{j} (p)|^2 \rangle
  = \int_{\mathcal{A}_{j}} dx \, d x^{\prime}  d x^{\prime \prime}
  dx^{\prime \prime \prime} \,
  Q ( x, x^{\prime}) \nonumber \\
  \times h ( x - x^{\prime \prime})
  h^\ast ( x^{\prime} - x^{\prime \prime \prime})
  \alpha_{j,p} ( x^{\prime \prime})
  \alpha_{j,p}^\ast (x^{\prime \prime \prime}).
\end{eqnarray}
Here $\langle \cdot \rangle $ denotes the averaging performed by the
detection and $ Q (x, x^{\prime} ) = \langle \Phi (x) \Phi^\ast
(x^{\prime}) \rangle$ is the mutual coherence function, which fully
characterizes the source properties (up to second order).  The
combined factor $\alpha_{j,p} ( x ) = A_{j} (x) \exp(ik x p/f)
$ condenses the lens focusing. Without loss of generality we can
assume a smooth Gaussian approximation of this aperture profile, so
that   $A_{j} ( x ) = \exp [ -\case{1}{4} (x - x_{j})^2/ (\Delta x)^2 ]$,
where $x_{j}$ is the position of the center of the $j$th microlens and
$(\Delta x)$ is the Gaussian width, which can be considered identical
for them all.

To show in a very simple manner the operating principle of these
devices, let $\phi(x)$ represent the wave front under consideration.
The mutual coherence function then reads $ Q(x, x^{\prime}) = \exp\{ i
k [ \phi (x^{\prime} ) - \phi(x) ] \}$. If $\phi (x)$ changes slowly
at each microlens, it can be approximated by a linear expansion near
the central point, so that $ \phi(x) \simeq \phi(x_j) + (x - x_j)
\phi^{\prime}(x_j) \ldots$, where $\phi^{\prime}(x_{j})$ is the
derivative evaluated at $x_{j}$. Keeping the Gaussian approximation, 
the signal due to each microlens can written, after a simple integration, as
\begin{equation}
  \label{geometrical limit}
  S_{j,p} \simeq \exp \{ - 2 k^{2 }(\Delta x)^2 \,
  [ p/f -  \phi^{\prime} (x_{j})  ]^{2}  \} \, .
\end{equation}
The limit of geometrical optics can be characterized by the condition
$k^2 (\Delta x)^2 \rightarrow \infty$. By employing the identity $
\lim_{\alpha \rightarrow \infty } \sqrt{\pi / \alpha} \exp ( - \alpha
t^2 ) = \delta(t)$, we immediately get
\begin{equation}
  \label{limit HS}
  S_{j,p} \simeq \delta (   p/f - \phi^{\prime} (x_{j})  )
\end{equation}
whose interpretation is quite convincing: the detection can be viewed
as a local tilting of the wave front. This also shows that the
displacement of the focal spot can be used to find the gradient of the
surface $\phi (x)$, whence we can compute the  wave front from a
direct integration (or other adapted numerical schemes).

\begin{figure}
 \centerline{\includegraphics[height=5cm]{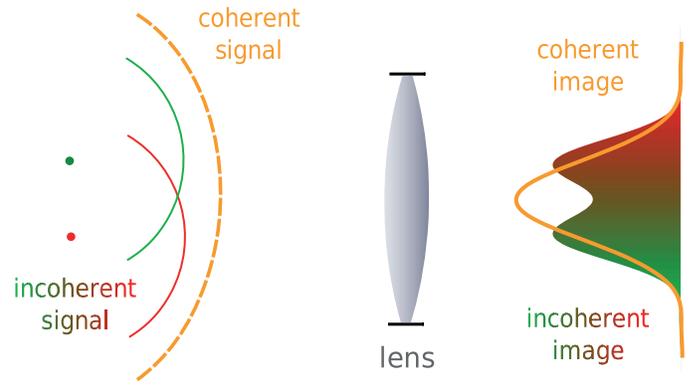}}
  \caption{Coherent (orange) and incoherent (red/green) images of two
    point sources separated by the Rayleigh distance. Notice how the
    recorded image varies with respect to the coherence properties of
    the signal. \label{fig:image}}
\end{figure}

However, this is an approximation: the scheme depends also on the
coherence of the signal. Loosely speaking, coherence expresses a
mutual relation between two or more sources.  Consider, for example,
the imaging of two distant pointlike sources described,
in the far field, by the plane waves $\exp(i\mathbf{\mathbf{k}}_1
\cdot \mathbf{x})$ and $\exp(i\mathbf{\mathbf{k}}_2 \cdot
\mathbf{x})$ (see Fig.~\ref{fig:image}). Two limiting cases can be
distinguished: for an incoherent superposition, the detector will
record two spots indicating the two respective directions, while for
coherent signals, the two plane waves will superpose resulting in a
single wave with a modulated amplitude $2 \cos [ (\mathbf{k}_1 -
\mathbf{k}_2) \cdot \mathbf{x} /2] \exp [ i (\mathbf{k}_1 +
\mathbf{k}_2) \cdot \mathbf{x} /2]$. In general, coherences in the
signal will have a significant effect on the detected intensity
profile. For example, when the two sources are separated by the
Rayleigh distance the detected image consists of a single blurred
point. In this particular case, two coherent point sources are not
resolved by a single aperture, while incoherent sources at the same
positions are just resolved.  In practice, the situation is even
trickier: any realistic signal is always generated by a superposition
of many plane waves, which may only be partially coherent.

Though the role of the coherence in image processing is clear,
standard optical arguments always assume coherent or incoherent
signals~\cite{Goodman:2005}.  Partial coherence is mostly assumed in the
context of interferometry and manifests by a reduction of the visibility.
In fact, the celebrated van Cittert-Zernike theorem states the way
in which coherence changes during the propagation.

Coherence is the main obstacle for the geometrical interpretation of
the signal in a Shack-Hartmann sensor. Indeed, when registering the
position and direction, there is still the uncertainty whether a
single ray or several interfering rays were registered.  This problem
is solvable only at the level of the mutual coherence function. In
view of these arguments the role of coherence seems to be
underestimated in current data processing and it might provide new
schemes. This is why we wish to bring into prominence the idea that
the performance of optical scanning devices can be pushed up to the
ultimate limits by resorting to tomographic reconstruction methods.

\textit{Quantum formulation.---} We now make the important observation
that the central equation (\ref{projection}) can be recast as
\begin{equation}
  \label{quantum formulation}
  S_{j,p}  = \langle  \alpha_{j,p} | U^{\dagger}  Q  U
  | \alpha_{j,p} \rangle ,
\end{equation}
where we have introduced an obvious Dirac notation. The unitary
transformation $U$ accounts for the propagator. For example, employing
the $x$ representation, we have that $h(x - x^\prime) = \langle x| U |
x^\prime \rangle $, which gives a direct meaning to $U$.
In addition, $Q(x, x^\prime) = \langle x| Q |x^{\prime} \rangle$ plays
the role of a density matrix and $\alpha_{j,p} (x) = \langle x|
\alpha_{j,p} \rangle $.  Equation~(\ref{quantum formulation}) may
appear as a simple reformulation of the problem; however, it is the
basis for all our results: it can be seen as the optical
interpretation of the quantum projection postulate.  The scanning
apparatus provides thus the same kind of information about the mutual
coherence function as an ordinary measurement does about the state of a
quantum system. 

Note that special cases of (\ref{quantum formulation}) are the
position intensity scan $I(x) = \langle x| Q | x \rangle $ and the
spectral (angular) intensity scan $ I (p) = \langle p | Q | p \rangle
$, where $| p \rangle$ are the momentum (angular) states obtained from
$ | x \rangle$ by Fourier transformation. Therefore, the intensity
$I(x)$ is a projection onto a precisely defined position, while the
spectral intensity is a projection along the direction of the incoming
wave. These intensities are measured by all optical devices with
imaging capabilities, such as telescopes and photographic cameras.

To go beyond the standard image processing, we propose to reinterpret
the operation of these devices as a generalized measurement of
noncommuting variables.  Indeed, position and momentum in the
transverse plane cannot be measured simultaneously with arbitrary
precision. In a quantum language, this represents a projection into a
minimum uncertainty (squeezed) state, determined by the effective
width of the $\Delta x$ detection delimited by the microlens
aperture. In terms of optics, the conjugate variable can be attributed
to the impulse $ p_x \simeq k p /f$, and fulfills the Heisenberg
uncertainty principle $\Delta x  \Delta p_{x}  \ge \hbar/2$.

Simultaneous detection of noncommuting variables (such as position
$x$ and momentum $p_{x}$) is well known in quantum optics. This
issue was addressed by Arthurs and Kelly~\cite{AK:1965} and can be
properly formulated as a quantum estimation
problem~\cite{Stig:1992,Raymer:1994}: one looks for the registered
classical variables, let us call them $X$ and $P_{x}$, that brings
information about original quantum observables $x$ and $p_{x}$.
These new variables obey an uncertainty relation in which the
lower bound is twice as large as the original Heisenberg
uncertainty $\Delta X \Delta P_{x}  \ge \hbar$.

The simultaneous measurement of $x$ and $p_{x}$ can be also modeled as
the measurement of the non-Hermitian operator $A = x + i
p_x$. Denoting the eigenstates of this operator as $| \alpha \rangle$,
$\alpha$ being the complex number with the real part corresponding to
the observable $ x $ and the imaginary part corresponding to $p_{x}$,
the probability distribution for their simultaneous detection is the
projection $ P_H (\alpha) = \langle \alpha | \varrho| \alpha \rangle$, 
also known as the Husimi function~\cite{Husimi:1940} of the quantum
state $\varrho$. Since $P_H$ is fully equivalent to $\varrho$, the
projections $|\alpha\rangle$ are informationally complete.  In
practice, however, $P_H$ can be sampled only by a finite number of
detections $\alpha$ and the tools of quantum state
estimation~\cite{PRL,lvovsky1,lvovsky2,bellini,grangier,polzik}
are especially germane to reconstruct $\varrho$ from experimental
data.

The strategy we suggest is straightforward:

(i) The signal is measured with a wave front sensor consisting of a
position sensitive detector placed in the focal plane of an array of
microlenses. Alternatively a single lens can be moved with respect to
the signal.

(ii) To each pixel (of angular coordinate $p$) in the focal plane of
the $j$th subaperture we associate a projection $|\alpha_{j,p}
\rangle$ performed on the signal beam with the expectation value
$S_{j,p}$. As we have shown above, in this way the signal
reconstruction is converted to a quantum state reconstruction problem.

(iii) The mutual coherence function $Q$ of the signal is reconstructed
by inverting the relation~\eqref{quantum formulation} with the
constraint $Q \ge 0$. For example, the maximum likely mutual coherence
matrix is obtained by iteratively solving the operator equation $RQ =
GQ$~\cite{NJP}, where
\begin{eqnarray}
  \label{eq:RG}
  R & = & \sum_{j,p} \frac{s_{j,p}}{S_{j,p}}
  |\alpha_{j,p} \rangle \langle \alpha_{j,p}| \, , 
  \nonumber \\
  G & = & \frac{\sum_{j,p} s_{j,p}}{\sum_{j,p}S_{j,p}}
  \sum_{j,p} |\alpha_{j,p} \rangle \langle \alpha_{j,p}|  
\end{eqnarray}
are positive semidefinite matrices, and $s_{j,p}$ are measured (noisy)
data.

Once $Q$ is known, any information of interest, such as, e.g.,
position or angular intensity scans in any transverse plane, coherence
properties, etc., can be obtained from $Q$ by postprocessing.  Note
also that the quantum aspects of the problem are manifested by the
positive semidefiniteness constraint $Q \ge 0$.

\begin{figure}[b]
   \centerline{\includegraphics[height=3cm]{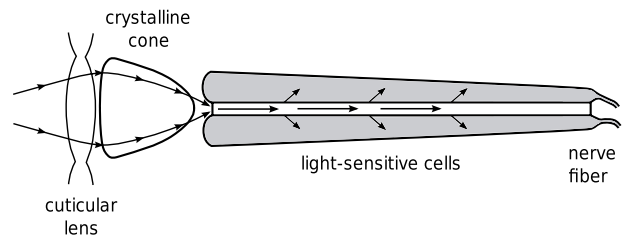}}

\caption{Schematic drawing of a cut through an ommatidium of a fruit fly.
The process of coupling the signal into the the light sensitive cell 
and the observed position,where  the light is absorbed provide,
in principle, information about the angular spectrum of the incident light. 
In this respect each ommatidium resembles a subaperture of a Hartmann-Shack sensor.
\label{fig:fly}}
\end{figure}

\textit{Discussion.---}
The scanning devices of the type discussed above are not rare in
Nature. For example, the compound eyes of insects~\cite{compound}
consist of thousands of identical units (ommatidia) and bear some
similarity with a Hartmann-Shack sensor (see Fig.~\ref{fig:fly}). As
we have just argued, such a detector can, in principle, be used for
reconstructing the mutual coherence function of the observed signal.
This may have intriguing consequences for biology.  For example, it is
a commonly-accepted hypothesis that flies are short-sighted because
the compound eye lacks the ability to accommodate and therefore to
focus at different distances.  Things may not be that simple: since
the mutual coherence contains full information, focusing (or any other
optical transformation) can be done by postprocessing the registered
data and sharp images can be obtained without any need for optical
focusing.

\begin{figure}
  \begin{tabular}{c}
    \includegraphics[width=0.9\columnwidth]{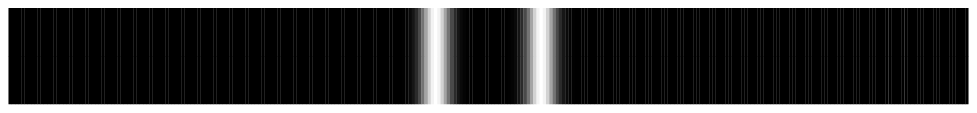}\\
    \includegraphics[width=0.9\columnwidth]{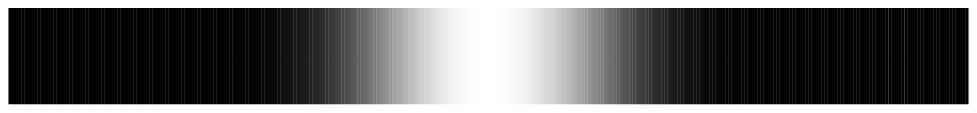}\\
    \includegraphics[width=0.9\columnwidth]{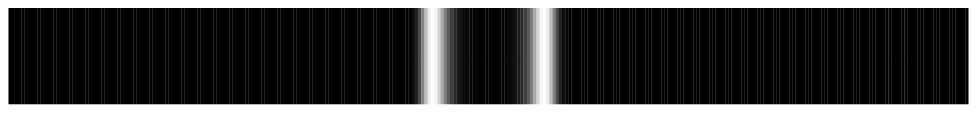}\\
  \end{tabular}
  \caption{Simulated reconstruction of a scanned optical signal.
    Panels from top to bottom: the simulated object, the blurred image
    registered by a short-sighted optical system, and the intensity
    distribution at the object plane as reconstructed from data
    provided by a scanning device. \label{fig:focus}}
\end{figure}

This can be clearly seen in Fig.~\ref{fig:focus}, which presents
an object consisting of a pair of spatially-separated bright
spots. For a particular choice of the parameters, a single blurred
spot is registered by a detector placed at the image plane of a
single lens. A Hartmann-Shack device with just 10 subapertures
fitted with 10-pixel detectors provides enough information to get
a faithful reconstruction of the original object. We leave as an
open question whether or not Nature actually takes advantage of
the physics behind these scanning devices.

As the second example, where the coherence could be challenging are
recent experimental results for temperature deviations of the cosmic
microwave background. The anisotropy is mapped as spots on
the sphere~\cite{CMB}, representing the distribution of directions of
the incoming radiation. To get access to the position distribution, the
detector has to be moved and, in principle, such a scanning sensor
then brings information about the position and direction
simultaneously. When the aperture is moving, it scans the
field repeatedly at different positions, denoted here by the index
$j$.  Consequently, the registered signal expressed in a fixed
reference frame reads $S_{j,p} = \langle \alpha_{p} |U_{j}^{\dagger} Q
U_{j} | \alpha_{p} \rangle$.

The ideal blackbody radiation in thermal equilibrium should be
homogeneous, isotropic and incoherent, as a consequence of
Bose-Einstein statistics. However, recent missions of the projects
COBE and WMAP have demonstrated convincingly that there is an
anisotropy in the distribution of cosmic microwave background
radiation. These deviations encode a wealth of information on the
properties of the Universe in its infancy. The objective of the Planck
mission is to measure these properties with unprecedented accuracy and
level of details~\cite{Planck}. This could be also an excellent chance
to investigate the subtle coherence properties of the relict
radiation. To our best knowledge, this question has not been posed
yet.  Though the coherence cannot be measured directly by standard
interferometric techniques, it could be inferred indirectly from the
image patterns generated by partially coherent sources. The methods of
quantum tomography are especially adequate for this task.

In summary, coherence can be used as an important source of
information, provided that the mutual coherence is properly sampled by
the detection. This possibility is not exploited in the standard
protocols of image processing. The proposed statistical inversion,
according to the receipts of quantum tomography, is the main result of
this Letter. Notice that the acquired mutual coherence function has
the extraordinary property of recording all views of an object and the
lack of resolution or accommodation could be more than compensated by
the remarkable possibility of postprocessing this complete
information, once it is accessible. This postprocessing is compatible
with the laws of Nature, and can surely be a source of inspiration for
the technology.

We acknowledge Hubert de Guise for a careful reading of this
manuscript.  This work was supported by the Czech Ministry of
Education, Project MSM6198959213, the Czech Ministry of Industry and
Trade, Project FR-TI1/364, and the Spanish Research Directorate, Grant
FIS2008-04356.


\begin{thebibliography}{10}

\bibitem{Tyson:1998}
R. K. Tyson,
{\em Principles of Adaptive Optics}
(Academic Press, Boston, 1998).

\bibitem{Roddier:1999}
F. Roddier,
{\em Adaptive Optics in Astronomy}
(Cambridge University Press, Cambridge, 1999).

\bibitem{Geary:1995}
J. M. Geary,
{\em Introduction to Wavefront Sensors}
(SPIE Press, Bellingham, 1995).

\bibitem{Buse:2000}
K. Buse and M. Luennemann
Phys. Rev. Lett. \textbf{85}, 3385 (2000)

\bibitem{Rehacek:2004}
{\em Quantum State  Estimation},
edited by M. G. A. Paris and J.
\v{R}eh\'{a}\v{c}ek,
Lect. Not. Phys. Vol.649
(Springer, Berlin, 2004).

\bibitem{Platt:2001}
B. C. Platt and R. S. Shack,
J. Refract. Surg. \textbf{17}, 573 (2001).

\bibitem{Goodman:2005}
J. W. Goodman,
{\em Introduction to Fourier Optics}
(Roberts, Greenwood Village, Colorado, 2005).


\bibitem{AK:1965}
E. Arthurs and J. L. Kelly Jr.,
Bell Syst. Tech. J. \textbf{44},  725 (1965).

\bibitem{Stig:1992}
S. Stenholm,
Ann. Phys. (NY)  \textbf{218}, 233 (1992).

\bibitem{Raymer:1994} M. G. Raymer, Am. J. Phys. \textbf{62},
986 (1994).

\bibitem{Husimi:1940}
K. Husimi,
Proc. Phys. Math. Soc. Japan \textbf{22}, 264 (1940).

\bibitem{PRL}
Z. Hradil, D. Mogilevtsev and J. \v{R}eh\'{a}\v{c}ek,
Phys. Rev. Lett. \textbf{96}, 230401 (2006).


\bibitem{lvovsky1}  
A.~I.~ Lvovsky, H.~Hansen, T.~Aichele, O.~Benson, J.Mlynek, and
S.~Schiller, 
Phys. Rev. Lett. \textbf{87}, 050402 (2001).

\bibitem{lvovsky2}
S.~A.~Babichev, J.~Appel, and A.~I.~Lvovsky, 
Phys. Rev. Lett. \textbf{92}, 193601 (2004).

\bibitem{bellini} 
A. Zavatta, S. Viciani, and M. Bellini, 
Science \textbf{306}, 660 (2004).

\bibitem{grangier}
A.~Ourjoumtsev, R.~Tualle-Brouri and P.~Grangier,
Phys. Rev. Lett. \textbf{96}, 213601 (2006).

\bibitem{polzik}
J. S. Neergaard-Nielsen, B. M. Nielsen, C. Hettich,
K. Moelmer and E. S. Polzik,
Phys. Rev. Lett. \textbf{97}, 083604 (2006).

\bibitem{NJP}
J. \v{R}eh\'{a}\v{c}ek, D. Mogilevtsev and Z. Hradil,
New J. Phys. \textbf{10}, 043022 (2008).

\bibitem{compound}
E. Warrant and D.-E. Nilsson (Eds.)
\textit{Invertebrate Vision}
(Cambridge University Press, Cambridge, 2004).

\bibitem{CMB}
G. Hinshaw \emph{et al},
Astrophys. J. Suppl S \textbf{180}, 225 (2009).

\bibitem{Planck}
See  http://www.esa.int/SPECIALS/Planck/index.html
\end{thebibliography}
\end{document}